\documentclass[sigconf]{acmart}
\usepackage{amssymb}
\usepackage{amsfonts}
\usepackage{algorithmic}
\usepackage{graphicx}
\usepackage{xcolor}
\usepackage{makecell}
\usepackage{booktabs} 
\usepackage{subfig}
\usepackage{amsmath}
\usepackage{accents}
\usepackage{statex}
\usepackage[normalem]{ulem}
\usepackage{enumitem}
\usepackage{multirow}
\usepackage[nomargin,inline,marginclue,draft]{fixme}
\usepackage{balance}
\usepackage{changepage}
\usepackage{bm}
\usepackage{hyperref}
\usepackage{setspace}
\usepackage{mathrsfs}
\usepackage{ulem}
\usepackage{verbatim}
\usepackage{diagbox}
\usepackage{pdftexcmds}
\usepackage{catchfile}
\usepackage{ifluatex}
\usepackage{ifplatform}
\usepackage{threeparttable}
\usepackage[normalem]{ulem}
\useunder{\uline}{\ul}{}
\usepackage{comment}

\usepackage[T1]{fontenc}
\usepackage{aecompl}

\newlength\savedwidth

\renewcommand{\authors}{}
\renewcommand{\shortauthors}{Xiaochong Lan et al.}

\author{Xiaochong Lan}
\affiliation{%
 \institution{Department of Electronic Engineering, BNRist,}
 \institution{Tsinghua University}
 \city{Beijing}
 \country{China}
}
\email{lanxc22@mails.tsinghua.edu.cn}

\author{Chen Gao$\dag$}
\affiliation{%
 \institution{Department of Electronic Engineering, BNRist,}
 \institution{Tsinghua University}
 \city{Beijing}
 \country{China}
}
\email{chgao96@gmail.com}

\author{Shiqi Wen}
\affiliation{%
 \institution{Meituan}
 \city{Beijing}
 \country{China}
}
\email{wenshiqi@meituan.com}

\author{Xiuqi Chen}
\affiliation{%
 \institution{Meituan}
 \city{Beijing}
 \country{China}
}
\email{chenxiuqi@meituan.com}

\author{Yingge Che}
\affiliation{%
 \institution{Meituan}
 \city{Beijing}
 \country{China}
}
\email{cheyingge@meituan.com}

\author{Han Zhang}
\affiliation{%
 \institution{Meituan}
 \city{Beijing}
 \country{China}
}
\email{zhanghan56@meituan.com}

\author{Huazhou Wei}
\affiliation{%
 \institution{Meituan}
 \city{Beijing}
 \country{China}
}
\email{weihuazhou@meituan.com}

\author{Hengliang Luo}
\affiliation{%
 \institution{Meituan}
 \city{Beijing}
 \country{China}
}
\email{luohengliang@meituan.com}

\author{Yong Li}
\affiliation{%
 \institution{Department of Electronic Engineering, BNRist,}
 \institution{Tsinghua University}
 \city{Beijing}
 \country{China}
}
\email{liyong07@tsinghua.edu.cn}

\copyrightyear{2023}
\acmYear{2023}
\setcopyright{rightsretained}
\acmConference[KDD '23]{Proceedings of the 29th ACM SIGKDD Conference on Knowledge Discovery and Data Mining}{August 6--10, 2023}{Long Beach, CA, USA}
\acmBooktitle{Proceedings of the 29th ACM SIGKDD Conference on Knowledge Discovery and Data Mining (KDD '23), August 6--10, 2023, Long Beach, CA, USA}
\acmDOI{10.1145/3580305.3599874}
\acmISBN{979-8-4007-0103-0/23/08}

\makeatletter
\gdef\@copyrightpermission{
  \begin{minipage}{0.3\columnwidth}
   \href{https://creativecommons.org/licenses/by/4.0/}{\includegraphics[width=0.90\textwidth]{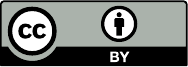}}
  \end{minipage}\hfill
  \begin{minipage}{0.7\columnwidth}
   \href{https://creativecommons.org/licenses/by/4.0/}{This work is licensed under a Creative Commons Attribution International 4.0 License.}
  \end{minipage}
  \vspace{5pt}
}
\makeatother

\begin{document}
\renewcommand{\shortauthors}{Xiaochong Lan et al.}
	\title{NEON: Living Needs Prediction System in Meituan}

\begin{abstract}
\footnotetext{$\dag$Chen Gao is the corresponding author (chgao96@gmail.com).}
Living needs refer to the various needs in human's daily lives for survival and well-being, including food, housing, entertainment, etc. At life service platforms that connect users to service providers, such as Meituan, the problem of living needs prediction is fundamental as it helps understand users and boost various downstream applications such as personalized recommendation. 
However, the problem has not been well explored and is faced with two critical challenges.
First, the needs are naturally connected to specific locations and times, suffering from complex impacts from the spatiotemporal context.
Second, there is a significant gap between users' actual living needs and their historical records on the platform. 
To address these two challenges, we design a system of living \textbf{NE}eds predicti\textbf{ON} named NEON, consisting of three phases: feature mining, feature fusion and multi-task prediction. 
In the feature mining phase, we carefully extract individual-level user features for spatiotemporal modeling, and aggregated-level behavioral features for enriching data,
which serve as the basis for addressing two challenges, respectively. 
Further, in the feature fusion phase, we propose a neural network that effectively fuses two parts of features into the user representation. Moreover, we design a multitask prediction phase, where the auxiliary task of needs-meeting way prediction can enhance the modeling of spatiotemporal context.
Extensive offline evaluations verify that our NEON system can effectively predict users' living needs. 
Furthermore, we deploy NEON into Meituan's algorithm engine and evaluate how it enhances the three downstream prediction applications, via large-scale online A/B testing. As a representative result, deploying our system leads to a 1.886\% increase \textit{w.r.t.} CTCVR in Meituan homepage recommendation.
The results demonstrate NEON's effectiveness in predicting fine-grained user needs, needs-meeting way, and potential needs, highlighting the immense application value of NEON.
\end{abstract}



\maketitle
	
\section{Introduction}\label{sec::intro}

\begin{figure}
\includegraphics[width=8cm]{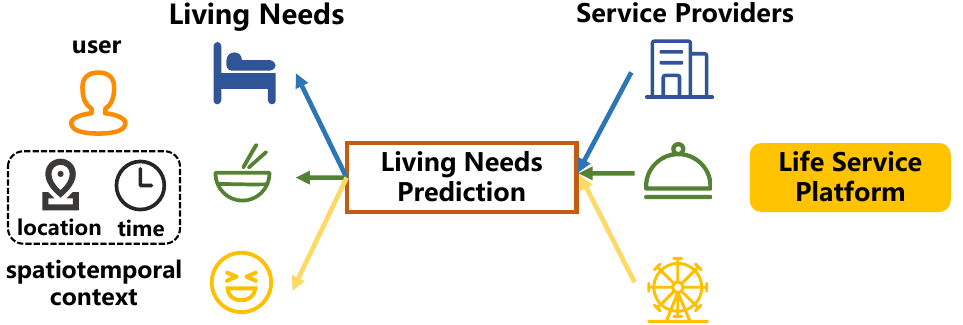}
\caption{Living needs prediction aims to predict the specific living need of a user given the spatiotemporal context. By predicting users' living needs, life service platforms can recommend life services that can fulfill these needs.}
\label{fig::nms}
\vspace{-0.8cm}
\end{figure}

\textit{Living needs} are the various needs generated by individuals in their daily routines for daily survival and well-being.
Typical living needs include necessities such as food, housing, and personal care as well as leisure activities such as entertainment, for which there exists a diverse array of life service providers. Meituan\footnote{https://www.metiuan.com} is a large platform connecting customers to life service providers, in which users can meet almost all kinds of living needs.
Unlike traditional information systems such as e-commerce websites, where users can only purchase products (\textit{i.e.}, meeting one kind of living needs), Meituan allows access to various living services, such as booking a hotel, ordering takeout
, etc. 
Moreover, an accurate understanding of users' needs can significantly improve user experience, for example, by enhancing the downstream recommendation tasks.
Generally speaking, generating a specific kind of living need is naturally accompanied by a specific time or location, such as ordering food delivery at the office.
Thus, the problem of \textit{living needs prediction} can be defined as predicting the type of needs of the target user's spatiotemporal context, as illustrated in Figure~\ref{fig::nms}.

For this new problem from the real-world scenario, there are two closely-related research topics, demand forecasting~\cite{lasek2016restaurant,seyedan2020predictive,ren2020demand,song2019review,hernandez2014survey} and spatiotemporal activity prediction~\cite{fan2019personalized,zheng2010collaborative,bhargava2015and,kipf2016semi,li2022disenhcn}. 
Demand forecasting is dedicated to predicting the quantity of a product or service that consumers will purchase. 
However, these works focus on the problem of aggregated demand prediction with times-series modeling, ignoring the needs of the specific user, time, and location.
As for spatiotemporal activity prediction,
the existing works focus mainly on either online activities such as App usage, or offline activities such as location visitation, while individuals' living needs can be fulfilled both in-store (offline) and via delivery (online), leading to a completely new problem. 

There exist two critical challenges for living needs prediction as follows.
\begin{itemize}[leftmargin=*]
    \vspace{-0.1cm}
    \item \textbf{The impact of spatiotemporal context is complex.} As discussed above, for the same user, living needs at different locations or times are totally different. For example, the user in Figure~\ref{fig::nms} eats food delivery at noon and watches a movie at night. Additionally, the lifestyles, \textit{i.e., the spatiotemporal pattern of living needs}, are extremely various for different users, further making it more difficult to model the spatiotemporal context.
    \item \textbf{There is a significant gap between users' actual living needs and their historical records on the platform. }
    Typically, users will face various kinds of situation in their life, and will generate multiple living needs. But they may only choose to satisfy one or a few of them on the platform, leading to a significant gap between their actual living needs and their historical records. We refer to the needs that can not be observed from historical records as \textit{potential needs}. The case here is different from that in most recommender systems, where the actual interests and the historical record generally do not differ greatly for users, leading to another critical challenge.
\end{itemize}
To address these challenges, in this work, we described our deployed NEON system (short for living \textbf{NE}eds predicti\textbf{ON}) in Meituan, which includes three phases: \textbf{feature mining}, \textbf{feature fusion}, and \textbf{multitask prediction}. First, in the feature mining phase, to address the first challenge, we carefully design the spatial and temporal features for individual-level users, and to address the second challenge, we extract the behavioral-pattern features for group-level users.
Second, in the feature fusion phase, we develop a feature-fusion neural network that combines internal preferences, impact from spatiotemporal context, and group behavior patterns to generate user representations, addressing both challenges.
Last, as the complement to the main task of living needs prediction, we introduce the auxiliary task of needs-meeting way prediction to enhance the model's learning of spatiotemporal context, further addressing the first challenge. 

The proposed NEON system plays a critical role in Meituan's recommendation engine with various downstream applications, including homepage recommendation, \textit{Guess-you-like} page recommendation, and message pop-up recommendation, which requires the different-aspect ability of living needs prediction. 
After the deployment of the proposed system, we obtain stable and significant gains in three applications, providing strong real-world evidence for NEON's effectiveness from different perspectives.

The contribution of this work can be summarized as follows.
\begin{itemize}[leftmargin=*]
    \item To the best of our knowledge, we take the first step to study the problem of living needs prediction, which is a critical problem in real-world life service platforms but has not been well explored.
    \item We proposed the NEON system, which includes three phases of feature mining, feature fusion layer, and multitask prediction, which well addresses the two challenges, the complex impact of spatiotemporal context and missing behavioral data. 
    \item We deploy NEON in Meituan's recommendation engine and conduct large-scale online A/B tests on three typical downstream applications, along with extensive offline evaluations. Offline experimental results verify that NEON can accurately predict users' living needs. The downstream evaluations strongly confirm NEON's high application value, with significant performance improvement in three downstream applications, among which a representative result is a 1.886\% increase \textit{w.r.t.} CTCVR for Meituan homepage recommendation.
\end{itemize}

\section{Problem Statement}\label{sec::profdef}
As we discussed above, in users' daily life, they generate various living needs, such as eating, accommodation, entertainment, beauty, etc. These needs can be fulfilled by life service providers in the city, which can be accessed through platforms connecting life service providers to customers. To enhance the user experience, it's crucial for these platforms to accurately predict users' needs and recommend appropriate services. This leads to the problem of living needs prediction. 

As defined in the introduction, living needs prediction aims to predict the specific living needs of a user given the spatiotemporal context in which they are located (in the following we also refer to this as given the \textit{user scene}). To clearly define the problem, with the help of experts, we divide all the living needs that users can satisfy on the platform into ten categories, shown in Table~\ref{tab::needs}. We use $\mathcal{N}$ as the symbol for the set of all living needs. In response to these needs, all the life services on Meituan are also divided into 10 categories. The problem can be defined as follows.

\begin{table}

\caption{10 types of living needs that can be satisfied in Meituan}
\vspace{-0.3cm}
\label{tab::needs}

\begin{tabular}{|c|l|}

\hline 
& \fontfamily{ppl}\selectfont Ordering food delivery, Eating in a restaurant, \\

 Living & \fontfamily{ppl}\selectfont Booking a hotel, Buying medicine, \\

 Needs & \fontfamily{ppl}\selectfont Specialty shopping online, Hair-dressing, \\

  &  \fontfamily{ppl}\selectfont Grocery shopping online, Beauty, \\

  & \fontfamily{ppl}\selectfont Tourism and Entertainment \\

\hline

\end{tabular}

\end{table}

\noindent \textbf{Input:}  A dataset $\mathcal{O}^+$ of real-world life service consumption records that reflect users' living needs. Each instance in the dataset tells the kind of life service a specific user purchases, which indicates the specific living need $n$ ($n\in \mathcal{N}$) of the user in a specific user scene $i$.

\noindent \textbf{Output:} A model to estimate the probability that a user will generate the living need $n$ in user scene $i$, formulated as $f(i,n|\mathcal{O}^{+})$. Here $f(\cdot)$ denotes the function that the model aims to learn.

\section{Our NEON System}\label{sec::method}

\begin{figure}
\vspace{-0.2cm}
\includegraphics[width=8cm]{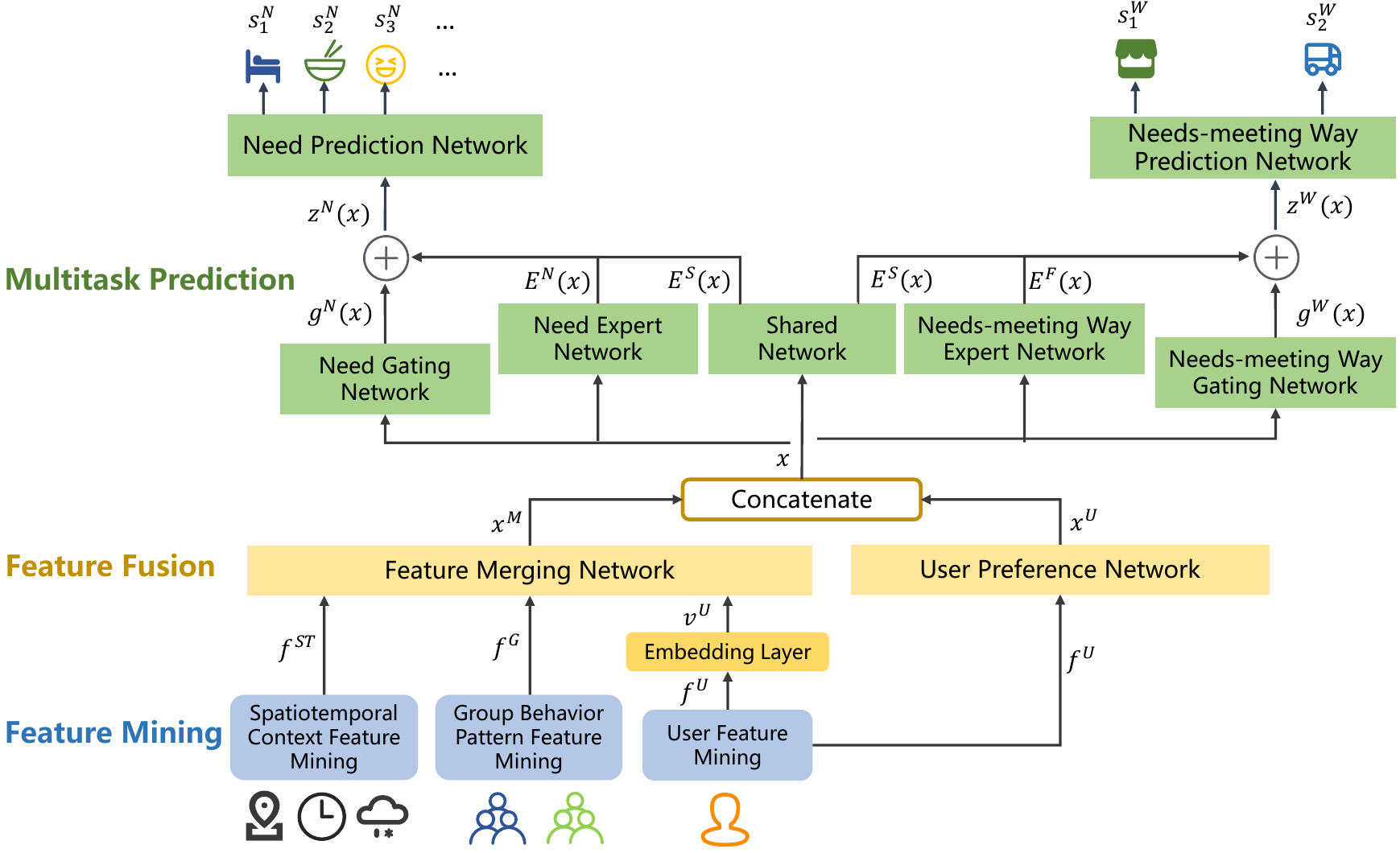}
\vspace{-0.2cm}
\caption{Illustration of living needs prediction system NEON.}
\vspace{-0.5cm}
\label{fig::model}
\end{figure}

To address the challenges mentioned in the introduction, we develop the NEON system made up of three phases: feature mining, feature fusion layer, and multitask prediction. 
First, in the feature mining phase, we address the first challenge by carefully designing spatiotemporal features for individual-level users and address the second challenge by extracting behavioral-pattern features for group-level users.
The feature fusion phase then employs a feature-fusion neural network to seamlessly integrate internal preferences, spatiotemporal context impact, and group behavior patterns to generate complete user representations, overcoming both challenges.
Last, in the multitask prediction stage, to enhance the model's understanding of spatiotemporal context, we introduce an auxiliary task of needs-meeting way prediction to the main goal of predicting living needs, providing additional support in addressing the first challenge. The deep feature fusion layer and the multi-task prediction parts of our system are illustrated in Figure~\ref{fig::model}.

\subsection{Feature Mining}\label{sec::FeatureMining}
First of all, we use features that directly reflect user traits, such as users' profile, their recent behavior sequence, and their historical behaviors, as inputs for the model.

 As mentioned above, for a specific user, his living needs are greatly affected by the spatial and temporal scenarios in which he is located. For example, \textit{on a rainy midday, a person at work probably has the need to order food delivery; but on a sunny noon, he/she may have another need of going out to eat in the restaurant}. This complexity and variability of human living needs, driven by the flux of time and space, pose a considerable challenge in accurately modeling the impact of the spatiotemporal context. In order to tackle this challenge, we incorporate spatiotemporal context features as an integral part of our system's input.

What's more, on the platform, users may have potential needs with sparse or even non-existent history records. For example, \textit{a person who never buys medicine on the platform may have a cold and need to buy medicine online one day.} Such potential needs are difficult for the model to grasp. To address the challenge of modeling potential needs, we introduce group behavior pattern features to help the model learn the potential living needs of users. 

Below we give a detailed description of the three categories of features.
\subsubsection{User Features}
This group of features includes user profiles and user history behavior sequences.
\begin{itemize}[leftmargin=*]
\item\textbf{User profiles $\bm{f}^U_p$.} The user's profile, including their age, gender, etc.
\item\textbf{User recent online behavior sequence $\bm{f}^U_{rb}$.} The sequence of items recently clicked by the user in the platform; the sequence of items recently ordered by the user in the platform.
\item\textbf{User aggregated historical online behavior $\bm{f}^U_{hb}$.} The percentage of times users buy each type of life service.
\item\textbf{User offline visitation record $\bm{f}^U_{ov}$.} The 50 most visited POIs (point of interest) by the user in the last six months; the 50 most visited AOIs (area of interest) by the user in the last six months. 
\end{itemize}
We concatenate all the mentioned features above to get a sparse user feature vector $f^U$, formulated as follows:
\begin{equation}
f^U=\left[f^U_p, f^U_{rb}, f^U_{hb}, f^U_{ov}\right].
\end{equation}
\subsubsection{Spatiotemporal Context Features}
Users' living needs are greatly affected by time, location, and other environmental factors. Thus, we introduce spatiotemporal context features as part of the input of our system to help our system model the complex impact of spatiotemporal context, which can be listed as follows.
\begin{itemize}[leftmargin=*]
\item\textbf{Time $\bm{f}^{ST}_t$.} Current time period. More than one time period feature of different granularity is applied, including hour, day, whether it is a holiday, etc.
\item\textbf{Location $\bm{f}^{ST}_l$.} The POI (point of interest) embedding of the user's real-time location; the AOI (area of interest) embedding of the user's real-time location; the city embedding of the user's real-time location. The location features are hourly real-time features.
\item\textbf{Weather $\bm{f}^{ST}_w$.} Weather information for the user's city or region, including wind, humidity, temperature, and weather type (sunny, rainy, snowy, etc.). Weather features are refined to hourly granularity.
\item\textbf{Travel state $\bm{f}^{ST}_{ts}$.} Information about whether the user is located in his/her resident city. Possible states include \textit{based in resident city}, \textit{about to travel}, and \textit{on travel}.
\end{itemize}
The dense spatiotemporal context feature vector $f^{ST}$ is created by concatenating all previously mentioned context features, formulated as follows:
\begin{equation}
f^{ST}=\left[f^{ST}_t, f^{ST}_l, f^{ST}_w, f^{ST}_{ts}\right].
\end{equation}
\subsubsection{Group Behavior Pattern Features}
We introduce group behavior pattern features to supplement the sparse individual behavior of users, in order to assist in identifying the potential living needs of individual users. 
\begin{itemize}[leftmargin=*]
\item\textbf{Group aggregated behavior $\bm{f}^G_{a}$.} We first segment users into groups based on their profiles. In each group, we get the group aggregated behavior by calculating the percentage of views, clicks, and purchases of each type of life service among all views, clicks, and purchases initiated by the group. For each user, the group aggregated behaviors of the groups the user belongs to are used as features. For example, a middle-aged person has group aggregated behaviors feature of middle-aged users and other groups he/she is in. 
\item\textbf{Popularity in the current time period $\bm{f}^G_{ct}$.} We cut all time into time periods according to different criteria, such as whether it is 
a holiday, if it is morning, noon, or night, etc. Then in each time period, we calculate the popularity of each type of life service by calculating the percentage of times the life service is viewed/clicked/purchased among all views/clicks/purchases happening in this time period. We determine the time periods in which the current time is located, and use the popularity of each type of life service in these time periods as a feature. For example, if the user opens the app on Christmas night, popularity on holiday and popularity at night of each kind of life service are set as features.
\item\textbf{Group behavior pattern in spatiotemporal context $\bm{f}^G_{st}$.} By discovering group preferences in different spatiotemporal contexts, we further capture more fine-grained group behavior patterns. We calculate the percentage of views/clicks/purchases of each type of life service initiated by each group in each kind of spatiotemporal context. These fine-grained patterns are used as features of the model. For example, the group preference of middle-aged people at work at noon on working days are used to enrich the representations of every individual within this demographic in such spatiotemporal scenario.
\item\textbf{User behaviors augmented by inter-need correlation $\bm{f}^G_{ic}$.} There is an inherent association across different types of users' living needs. This association can be leveraged to improve prediction performance. For example, \textit{a user who frequently purchases hairdressing services may also be inclined to purchase beauty services.} We use the association rule mining algorithm to analyze the co-occurrence of different life service categories, filter out high-correlation relationships, and employ them to augment user behavior as input features.
\end{itemize}
We combine all previously mentioned group behavior pattern features to generate the dense group behavior pattern feature vector $f^{G}$, which is formulated as follows:
\begin{equation}
f^{G}=\left[f^{G}_a, f^{G}_{ct}, f^{G}_{st}, f^{G}_{ic}\right].
\end{equation}

\subsection{Feature Fusion Layer}\label{sec::FeatureIntegration}
As mentioned in Section~\ref{sec::profdef}, we refer to a user in a specific spatiotemporal context as a user scene. For a user scene $i$, after feature mining, we have dense spatiotemporal features $f^{ST}_i$, dense group pattern features $f^G_i$, and sparse user features $f^U_i$. For brevity of presentation, we omit the subscript $i$ in some of the expressions below. We designed a feature fusion layer to integrate these features into the input of the subsequent prediction module.

We first set up an embedding layer, which processes the high-dimensional sparse user feature vector $f^U$ into a low-dimensional dense vector $v^U$.  To address the challenges of complex impact of spatiotemporal context and users' potential needs, we mine spatiotemporal features and group behavior pattern features in the feature mining phase, respectively. With these features as input, we use a feature merging network to model the interaction between spatiotemporal contexts, group behavior patterns, and  users as follows,
\begin{equation}
x^M=h^M\left(\left[f^{ST}, f^G, v^U\right]\right),
\end{equation}
where $\left[ \cdot \right]$ denotes concatenation operation. Here $h^M$ is the feature merging network, which merges three information sources of spatiotemporal contexts, group behavior patterns, and user preference into a fusion representation $x^M$.

Moreover, users have their own internal characteristics that are independent of the spatiotemporal scene they are in and the group they belong to. To model the internal characteristics of users, we generate a representation as follows,
\begin{equation}
x^U=h^U\left(f^U\right),
\end{equation}
where $h^U$ denotes the user preference network that turns raw user features into dense user preference representation. We then concatenate the two parts of representations into the full representation of the user scene:
\begin{equation}
x=\left[x^M, x^U\right].
\end{equation}
In brief, we design a feature fusion layer to tackle both challenges by considering the influence of spatiotemporal context, incorporating group behavior patterns, as well as extracting individual preferences.

\subsection{Multitask Prediction}\label{sec::Model}
We further design a prediction module which takes user scene representation as input to predict users' living needs. The module is tasked with two objectives: \textit{fine-grained need prediction} and \textit{needs-meeting way prediction}. Fine-grained need prediction is to predict the specific living need of the user. Neets-meeting way prediction is to predict the preferred way of the user to meet their needs. 

Specifically, among the ten kinds of needs which we mentioned in the problem formulation, there are two ways to satisfy the needs: in-store and via-delivery. In other words, consumers can choose to satisfy their needs by visiting a physical store or by ordering online and then receiving goods via delivery. Each type of the 10 needs can be classified into one of two categories, in-store needs or via-delivery needs. We show the classification in Table~\ref{tab::nc}. Actually, needs-meeting way prediction is to predict whether the preferred way of the user to meet their needs is in-store or via-delivery.
\begin{table}[]
\vspace{-0.3cm}
\caption{The classification of the 10 living needs that can be satisfied on Meituan}
\label{tab::nc}
\begin{tabular}{|l|l|}
\hline Living needs that& \fontfamily{ppl}\selectfont Specialty shopping online,\\
 can be satisfied&  \fontfamily{ppl}\selectfont Grocecy shopping online, \\
via delivery& \fontfamily{ppl}\selectfont Ordering food delivery, Buying medicine \\
\hline Living needs that& \fontfamily{ppl}\selectfont Eating in a restaurant, Hotel, \\
can be satisfied& \fontfamily{ppl}\selectfont Hair-dressing, Beauty,  \\
in store& \fontfamily{ppl}\selectfont Tourism, Entertainment \\
\hline
\end{tabular}
\vspace{-0.6cm}
\end{table}

Users' preferences for needs-meeting ways are strongly affected by the spatiotemporal context. For example, \textit{a person at work during lunchtime on a weekday is more likely to have the need to order food delivery (via delivery), while the same person in a shopping district on a weekend evening is more likely to visit a store for a meal (in-store)}. With this in mind, we include the needs-meeting way prediction task which is jointed trained with the main task of need prediction to enhance the model's ability to learn spatiotemporal context information.
Next, we describe how we get the prediction results of the two tasks. 
We use $y^W$ and $y^N$ to denote the prediction result of needs-meeting way and specific need. $y^k, k \in \{W, N\}$ can be generated as follows,
\begin{equation}
 \begin{aligned}
y^k &= t^k(z^k),\\
\text{where }z^k &= g^k(x)_0E^k(x)+g^k(x)_1E^S(x).
\end{aligned}
\end{equation}
Here $t^k$ is the prediction neural network for task $k$. There are a variety of choices in the specific structure of the neural network. In Section~\ref{sec::implementation}, we will state our specific choice. To avoid verbosity, we will use \textit{network} to replace ~\textit{neural network} in the following text. The output of $t^N$, $y^N$, is the scores of ten types of living needs, and the output of $t^W$, $y^W$, is the scores of in-store and via-delivery needs-meeting ways. We use $s^N_{im}$ to denote the score of need $m$ for user scene $i$, and use $s^W_{in}$ to denote the score of needs-meeting way $n$ for user scene $i$. $E^k$ is the expert network~\cite{ma2018modeling,tang2020progressive} for task $k$. $E^S$ is the shared network between the two tasks. $E^S$ is responsible for generating general representations that are common to both tasks, while $E^k$ is responsible for learning task-specific representations that are more fine-tuned to the specific task $k$. The gating network $g_k$ determines what proportion of information input each task's prediction network receives from the shared network and the expert network. We formulate the gating network as follows,
\begin{equation}
g^k(x)=\operatorname{Softmax}(W_k x),
\end{equation}
where $W_k \in \mathbb{R}^{2 \times d}$ is trainable weights for task $k$. The gating network takes $x$ as input, and outputs the relative importance of the shared and task-specific representations for a given tasking, allowing the model to selectively attend to the most relevant information and improve its performance.
In summary, to address the complexity of spatiotemporal context impact, we introduce an auxiliary task of needs-meeting way prediction which is jointly trained with the main task of fine-grained living needs prediction to enhance our system's learning of spatiotemporal context. The multitask prediction module in our system produces a score for each living need and needs-meeting way.
\subsection{Model Training}\label{sec::train}
In this section we describe how our system is trained. Corresponding to the two tasks, we design two parts of loss. We design need prediction loss taking into account the fact that the frequency of different needs arising in users' lives is different. For example, \textit{a user may need to order food delivery for lunch every workday, but rarely need to buy medicine}. In order to address the class imbalance issue for different living needs, we propose using a multi-class focal loss which can decrease the effect of needs with a high volume of training data on the final prediction loss. The need prediction loss can be formulated as follows,
\begin{align}
\text {Loss}_{\text{need}}&=-\sum_{i\in O}\left(\sum_{n=1}^{10}\left(1-q^N_{i n}\right)^\gamma \chi^N_{i n} \log \left(q^N_{i n}\right)\right),\\
\text{where } q^N_{in}&=\operatorname{Softmax}\left(s^N_{i n}\right)=\frac{e^{s^N_{i n}}}{\sum_n e^{s^N_{i n}}}.
\end{align}
Here $O$ is the training set, $s^N_{in}$ is the score of living need $n$ for user scene $i$, $\gamma$ is the hyperparameter which decides the importance of difficult samples, $\chi^N_{in}$ is 1 if $n$ is the ground truth need for user scene $i$, else it is 0.
For the needs-meeting way prediction task, we use BCE loss as prediction loss. We formulate it as follows,
\begin{align}
\text{Loss}_{\text{way}}&=-\sum_{i\in O}\left(\sum_{m=1}^{2}\chi^W_{im}
\log \left(q^W_{im})\right)\right),\\
\text{where } q^W_{im}&=\operatorname{Softmax}\left(s^W_{i m}\right)=\frac{e^{s^W_{i m}}}{\sum_m e^{s^W_{i m}}}.
\end{align}
Here $O$ is the training set, $s^W_{im}$ is the score of needs-meeting way $m$ (in store or via delivery )for user scene $i$.
$\chi^W_{im}$ is 1 if $m$ is the ground truth needs-meeting way for user scene $i$, else it is 0.
In our system, the feature integration module and multitask prediction module are trained end to end. The entire loss function is:
\begin{equation}
\text{Loss} = \lambda_1 \text{Loss}_{\text{need}}+\lambda_2 \text{Loss}_{\text{way}}.
\end{equation}
$\lambda_1$ and $\lambda_2$ are hyperparameters that control the importance of the two parts of loss.

\section{Offline Evaluation}\label{sec::experiments}
\subsection{Experimental Settings}
\subsubsection{Dataset}
We conduct an offline experiment on a real-world dataset at a billion-scale. The dataset comprises a sampling of all 2022 purchase records on the platform, based on the percentage of purchases for each type of life service. It consists of over 7 billion actual purchase records from 65 million users. The details of datasets
are provided in Appendix~\ref{sec::dataset}.
\subsubsection{Metrics} 
We design three metrics, namely Sort Accuracy (SA), Via-delivery Sort Accuracy (VDSA), and In-store Sort Accuracy (ISSA), to measure the performance of our system and baseline systems in predicting living needs. SA measures the overall accuracy of sorting living needs based on their scores. VDSA focuses on the accuracy of predicting needs that are satisfied via delivery, while ISSA focuses on the accuracy of predicting needs for in-store scenarios. Detailed definitions of these metrics are provided in Appendix~\ref{sec::metrics}.
\subsubsection{Baselines}
To illustrate the effectiveness of our system, we compare it with two baselines widely in actual production environments, including \textbf{DIN~\cite{zhou2018deep}}, \textbf{DNN~\cite{cheng2016wide}}, \textbf{DCN~\cite{wang2017deep}}, 
\textbf{ESMM~\cite{ma2018entire}},
and \textbf{MMOE~\cite{ma2018modeling}}. We will provide a detailed description of these baselines in Appendix~\ref{sec::baselines}.
\subsection{Overall Performance}
We test the performance of our proposed system and baselines on the living needs prediction task, and show the results in Table~\ref{tab::overallperformance}. We can have the following observations.
\begin{table}[]
\caption{Offline experimental performance of NEON and baselines.}
\vspace{-0.2cm}
\label{tab::overallperformance}
\begin{tabular}{cccc}
\hline
\textbf{Method} & \textbf{VDSA}   & \textbf{ISSA}   & \textbf{SA}     \\ \hline
DIN             & 0.9044          & 0.7467          & 0.8700          \\
DNN             & 0.9060          & 0.7500          & 0.8718          \\
DCN             & 0.9051 & 0.7466 & 0.8701\\
ESMM & 0.9080 & 0.7476 & 0.8708 \\
MMOE & 0.9097 & 0.7573 & 0.8757 \\
NEON      & \textbf{0.9175} & \textbf{0.8277} & \textbf{0.9070} \\ \hline
Improvement     & 0.86\%          & 9.30\%         & 3.57\%          \\ \hline
\vspace{-0.5cm}
\end{tabular}
\end{table}

\begin{table}[]
\caption{Performance of NEON with/without multitask prediction.}
\vspace{-0.2cm}
\label{tab::ablation}
\begin{tabular}{cccc}
\hline
 & \textbf{VDSA}   & \textbf{ISSA}   & \textbf{SA}     \\ \hline
w.o.           & 0.9089          & 0.8084          & 0.8968          \\
with             & 0.9175          & 0.8277          & 0.9070          \\ \hline
Improvement     & 0.95\%          & 2.39\%         & 1.14\%          \\ \hline
\vspace{-0.5cm}
\end{tabular}
\end{table}

\begin{itemize}[leftmargin=*]
\item{\textbf{Our system steadily outperforms all baselines on all metrics.}} The improvement of our system compared to the best baseline is 0.86\%, 9.30\%, and 3.57\% \textit{w.r.t.} VDSA, ISSA, and SA, respectively. The significant performance gain confirms the effectiveness of our system on the living needs prediction task. Furthermore, such a significant improvement in the ability to predict living needs will result in a huge benefit in real-world production scenarios, which will be further confirmed through online evaluation.
\item{\textbf{Our system achieves greater improvement on ISSA.}} The task of predicting users' in-store living needs is relatively difficult, since the in-store consumption data is sparser, and the relationships between spatiotemporal context and in-store needs are more complex. On this task, all methods perform the worst, and our system outperforms baselines on a large margin, with an improvement of 9.30\% \textit{w.r.t.} ISSA. This further confirms our model's ability to tackle the complex impact of spatiotemporal context and discovering potential needs. 
\end{itemize}
\subsection{Ablation Study}
As mentioned in Section~\ref{sec::Model}, we introduce the task of needs-meeting way prediction to enhance the system's learning of spatiotemporal context. To study the effectiveness of the multitask prediction design, we remove it from our system to observe the impact of the design on the system performance. Specifically, we change the system structure by removing the needs-meeting way prediction network $t^W$ and taking the sum of $z^N$ and $z^W$ as the input of the need prediction network $t^N$, and test the performance of the changed system. The results are shown in Table~\ref{tab::ablation}.
The results show that our proposed system outperforms the system without multitask prediction design. Our system performs better \textit{w.r.t.} VDSA by 1.16\%, \textit{w.r.t.} ISSA by 2.39\%, and \textit{w.r.t.} SA by 1.14\%. The significant performance improvement confirms the validity of the multitask prediction design.
\section{Online Evaluation}\label{sec::online}
\begin{figure}
\includegraphics[width=8cm]{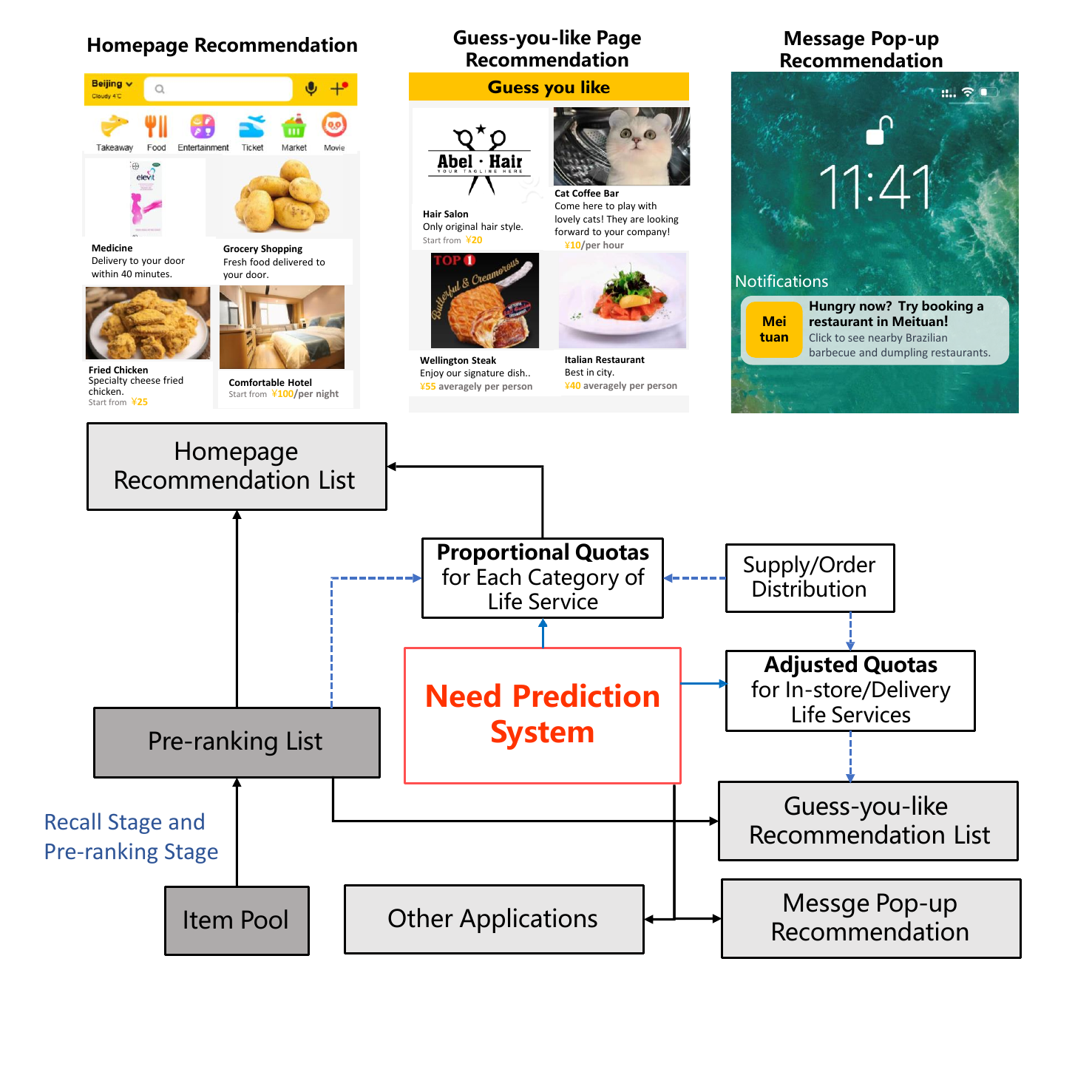}
\vspace{-0.8cm}
\caption{Online deployment of NEON in Meituan Recommendation Engine. In summary, NEON helps to generate the quotas of categories of life services in the recommendation list (homepage recommendation and \textit{guess-you-like} recommendation), or decide the one category to be recommended to the user (message pop-up recommendation).}
\label{fig::odp}
\vspace{-0.6cm}
\end{figure}
For life service platforms such as Meituan, understanding and predicting users' living needs are important in all business scenarios. In this section, we evaluate the performance of three downstream recommendation tasks when deploying NEON into Meituan’s recommender engine, and the illustration of deployment is shown in Figure~\ref{fig::odp}. Specifically, the three typical applications are homepage recommendation, \textit{Guess-you-like} recommendation, and message pop-up recommendation. The performance of NEON on these applications reflects its effectiveness on fine-grained need prediction, needs-meeting way prediction, and potential need prediction, respectively.
We elaborate on the online testing of three tasks one by one as follows.
\subsection{Homepage Recommendation}
In this section, we evaluate the performance of NEON when deployed to homepage recommendation, which reflects its ability on fine-grained living need prediction.
\subsubsection{Deployment Scenario}
Typically, users open Meituan mobile APP to meet their certain living need. In the overall recommendation list on homepage offered by Meituan, users probably only focus on the items which belong to the type of life service that can fulfill their living needs, and choose one item from the items. Whether an item is in the category of life service that the user \textit{need} is at least as important as whether the user \textit{likes} the item. To emphasize the importance of recommending \textit{needed} items, the engine follows a two-step approach to generate the final recommendation list from the pre-ranking result. For a user scene, it first decides the quotas for all the ten categories of life services and then generates the final list according to the quotas and the prediction scores of the items. The supply quotas are generated based on the scores our system outputs along with other criteria.

In short, our NEON system is used to generate the \textit{quotas} of different kinds of life services in the homepage recommendation list. 
\subsubsection{Experiment Setting}
We will first give a detailed description of how our system is used in generating the quotas for each category of life services. At first, we recall local life service items into the recall pool using various strategies, such as popularity and collaborative filtering. Then the engine outputs a preliminary recommendation list based on the recall pool, called pre-ranking list. After that, we take the Softmax normalized scores output by our system as the proportional quotas for each category of life service. We further adjust the quotas taking the proportion of each category in the pre-ranking list, supply distribution by category, and order distribution by category into account. The generated quotas are the proportion of each category in the lists received by users. The lists are generated considering both quotas of categories and prediction scores of items.

We compare the homepage recommendation performance of our whole recommendation engine with and without our system through online A/B tests. The tests last for one week and involved around 4.5 million users.

In our tests, the users are randomly divided into two buckets of similar size and assigned different methods for calculating quotas for each category. Specifically, for the first group, we use the method described previously, while for the second group, we generate quotas based only on the proportion of each category in the pre-ranking list, supply distribution by category, and order distribution by category. We maintain consistency across all other modules to ensure a fair comparison. 

For metrics, we use \textit{Click Through Rate} (CTR), \textit{Conversion Rate} (CVR), \textit{Click to Conversion Rate} (CTCVR) to measure the quality of the final recommendation list, which are widely-used measurements~\cite{ma2018entire,wen2020entire,liu2020autofis}. 
\begin{table}[]
\caption{Results of A/B tests on homepage recommendation}
\vspace{-0.2cm}
\label{tab::online1overall}
\begin{tabular}{cccc}
\hline
Metric & w/o NEON & with NEON & Improvement \\ \hline
CTR    & 3.0601$\times 10^{-2}$  & 3.0672$\times 10^{-2}$   & +0.230\%     \\
CVR    & 1.1497$\times 10^{-1}$ & 1.1687$\times 10^{-1}$  & +1.652\%     \\ 
CTCVR  & 3.5183$\times 10^{-3}$   & 3.5846$\times 10^{-3}$    & +1.886\%     \\  \hline
\end{tabular}
\vspace{-0.5cm}
\end{table}
\subsubsection{Performance} 
The results of our A/B tests are shown in Table~\ref{tab::online1overall}. From the results, we can have the following observations: 
\begin{itemize}[leftmargin=*]
\item{There is a significant improvement with respect to all metrics.} The increase \textit{w.r.t.} CTR and CVR are 0.230\% and 1.653\%, respectively, which is a notable improvement. 
\item{The increase \textit{w.r.t.} CTCVR is 1.886\%.} CTCVR indicates how likely users are to purchase the recommended items. Such improvement can result in a substantial rise in total consumption on the platform. 
\item{The rise \textit{w.r.t.} GTV-CC is 3.627\%.} The remarkable uplift demonstrates the outstanding capability of our system in predicting users' living needs that has no historical record. 
\end{itemize}

We further calculate the \textit{Kullback-Leibler divergence}~\cite{kullback1951information} (KLD) between real user order distribution by category and the average proportional allocation given by the online engine with/without our system. Lower KLD indicates the quotas match better with real user order distribution, which can be regarded as real user needs distribution. The results of different time periods throughout the day are shown in Table~\ref{tab::KLD}. The time periods are separated based on the business characteristics of the platform during each hour. Similar hours are grouped within a single period.
\begin{table}[]
\caption{KLD of different time periods between real user order distribution and quotas given with/without NEON.}
\vspace{-0.2cm}
\label{tab::KLD}
\begin{tabular}{cccc}
\hline
\multicolumn{1}{l}{\multirow{2}{*}{Time Period}} & \multicolumn{2}{l}{Kullback-Leibler Divergence} & \multicolumn{1}{l}{\multirow{2}{*}{Improvement}} \\ \cline{2-3}
\multicolumn{1}{l}{}                             & w/o NEON               & with NEON            & \multicolumn{1}{l}{}                             \\ \hline
0-4   & 0.2578 & 0.2201 & -14.62\% \\
5-8   & 0.2237 & 0.1907 & -14.75\% \\
9-10  & 0.2179 & 0.1902 & -12.71\% \\
11-12 & 0.2408 & 0.2138 & -11.21\% \\
13-16 & 0.2323 & 0.2053 & -11.62\% \\
17-19 & 0.2353 & 0.2096 & -10.92\% \\
20-24 & 0.2333 & 0.1995 & -14.49\% \\ \hline
\end{tabular}
\vspace{-0.5cm}
\end{table}

From the results, it can be seen that in all time periods, the Kullback-Leibler Divergence between the actual user order distribution by category and the average proportional allocation generated by the online engine with NEON is significantly less than that without our NEON. The percentage of decrease (improvement) is 12.90\% on average. This indicates that the quotas produced by the online engine with our system are more aligned with the actual user consumption distribution, or the real-life user living needs, compared to the ones generated without our system. This can be regarded as evidence of our model's effectiveness in addressing the intricate impact of spatiotemporal context, and further confirms our system's strong ability for predicting fine-grained user needs.
\subsection{Guess-you-like Recommendation}
This section assesses NEON's performance in guess-you-like recommendation, highlighting its ability on needs-meeting way prediction.
\subsubsection{Deployment Scenario}
Meituan designs a \textit{Guess-you-like} page, which user may be guided to during their leisure time. Typically, users browse this page to satisfy their \textit{unnecessary} living needs, such as entertainment or beauty, etc. In this page, they purchase items they need and also \textit{like}. To provide more choices for users in the categories of life services they need, the recommendation list should have more items in those categories. In this page, we assume that users are more concerned with being recommended items they like. Therefore, we do not maintain the proportion of a category in the "Guess-you-like" recommendation list just because there is a slight possibility that it is essential, as users can use modules other than Guess-you-like such as the search button to find necessary life services. To ensure enough items are in the needed category, the online engine generates new quotas for Guess-you-like page based on the quotas for homepage recommendation list and the in store/via delivery score output by our system. 

In summary, the needs-meeting way prediction results of NEON are used to further adjust the \textit{quotas} of different life services in \textit{Guess-you-like} recommendation list.
\subsubsection{Experiment Setting}
We conduct online A/B tests involving about 7 million users over a period of two weeks. We randomly divide users into two buckets, each of which has a similar amount of users, and assign them different methods of generating the quotas of categories in the Guess-you-like page. Specifically, for the first bucket we adopt the same strategy as in the homepage recommendation. For the second bucket, we calculated the score of both needs-meeting ways, and turn higher the quotas of categories whose needs-meeting way gets a higher score.
As for metrics, we use CTR, CVR, CTCVR, \textit{Negative Feedback Rate for Unique Visitor} (NFR-UV), \textit {Negative Feedback Rate for Page View} (NFR-PV) to measure the quality of the recommendation list in the Guess-you-like page. NFR-UV and NFR-PV emphasize that the recommendation list should not contain items users dislike. 
\subsubsection{Performance}
Under the aforementioned settings, we conduct extensive online A/B experiments. The results are shown in Table~\ref{tab::online2overall}. We list our observations as follows:
\begin{table}[]
\caption{Overall results of A/B tests on \textit{Guess-you-like} page recommendation.}
\vspace{-0.2cm}
\label{tab::online2overall}
\begin{tabular}{cccc}
\hline
Metric & w/o NEON & with NEON & Improvement \\ \hline
CTR    & 9.5707$\times 10^{-2}$  & 9.5647$\times 10^{-2}$    & -0.063\% \\
CVR    & 1.3568$\times 10^{-1}$  & 1.3606$\times 10^{-1}$    & +0.280\%     \\
CTCVR  & 1.3014$\times 10^{-2}$  & 1.2985$\times 10^{-2}$    & +0.218\%    \\
NFR-UV & 1.0128$\times 10^{-4}$  & 1.0294$\times 10^{-4}$    & +1.605\%     \\
NFR-PV & 8.3283$\times 10^{-5}$  & 8.6726$\times 10^{-5}$    & +3.342\%     \\ \hline
\end{tabular}
\end{table}
\begin{itemize}[leftmargin=*]
\item{The performance gain \textit{w.r.t.} CVR, CTCVR, and OV are 0.280\%, 0.218\%, and 0.310\%, respectively.} Such improvement resulting from adjusting the quotas with the assistance of the needs-meeting way prediction module can lead to a significant increase in the consumption amount.
\item{The NFR-UV and NFR-PV decrease by 1.57\% and 3.31\% respectively, indicating that the recommendation engine is able to recommend fewer items that users dislike in the Guess-you-like page by adjusting the quotas with the aid of the needs-meeting way prediction module in our system.} 
\end{itemize}

We further calculate the performance increase on some popular categories of life services in Guess-you-like page. The results are shown in Table~\ref{tab::bus}. 
From the result, we can observe that:
\begin{itemize}[leftmargin=*]
\item{For all these categories there is average relative improvement of 3.801\% \textit{w.r.t.} CVR and 3.765\% \textit{w.r.t} CTCVR.} With such improvements, all these categories can enjoy a notable increase in order volume on Meituan platform. 
\item{Among all these categories, the rise on Kids category is the most significant, which is 8.723\% \textit{w.r.t.} CVR and 9.691\% \textit{w.r.t.} CTCVR.} 
\item{There are also slight decreases \textit{w.r.t.} CTR of 0.034\% for Food Delivery, -0.449\% for Beauty, and -0.595\% for Hotel.} These decreases are within the typical fluctuations observed in the market. 
\end{itemize}
The above results, which demonstrate the successful implementation of our system in guess-you-like recommendation, further illustrate the efficacy of our system, particularly in needs-meeting way prediction.

\begin{table}[]
\caption{The A/B tests results on several popular categories of life services in \textit{Guess-you-like} page.}
\label{tab::bus}
\begin{tabular}{ccccc}
\hline
Category                  & Metric & w/o NEON   & with NEON  & Improvement \\ \hline
\multirow{3}{*}{Food} & CTR    & 1.8298$\times 10^{-2}$  & 1.8292$\times 10^{-2}$  & -0.034\%     \\
                          & CVR    & 2.8841$\times 10^{-1}$ & 2.8987$\times 10^{-1}$ & +0.506\%     \\
                          & CTCVR  & 5.2774$\times 10^{-3}$ & 5.3023$\times 10^{-3}$ & +0.472\%     \\ \hline
\multirow{3}{*}{Beauty}   & CTR    & 1.8769$\times 10^{-2}$  & 1.8684$\times 10^{-1}$  & -0.449\%     \\
                          & CVR    & 1.2583$\times 10^{-3}$  & 1.2889$\times 10^{-3}$ & +2.427\%     \\
                          & CTCVR  & 2.3616$\times 10^{-4}$  & 2.4081$\times 10^{-4}$  & +1.967\%     \\ \hline
\multirow{3}{*}{Kids}     & CTR    & 1.7717$\times 10^{-2}$  & 1.7874$\times 10^{-2}$  & +0.890\%     \\
                          & CVR    & 7.8825$\times 10^{-3}$  & 8.5701$\times 10^{-3}$  & +8.723\%     \\
                          & CTCVR  & 1.3965$\times 10^{-1}$  & 1.5319$\times 10^{-4}$  & +9.691\%     \\ \hline
\multirow{3}{*}{Hotel}    & CTR    & 1.6567$\times 10^{-2}$  & 1.6469$\times 10^{-2}$ & -0.595\%     \\
                          & CVR    & 4.1454$\times 10^{-2}$  & 4.2923$\times 10^{-2}$  & +3.546\%     \\
                          & CTCVR  & 6.8678$\times 10^{-1}$  & 7.0690$\times 10^{-4}$  & +2.930\%     \\ \hline
\end{tabular}
\vspace{-0.3cm}
\end{table}
\subsection{Message Pop-up Recommendation}\label{sec::pnp}
In this section, we test our model's performance in message pop-up recommendation to observe its potential need prediction ability.
\subsubsection{Deployment Scenario}
In the case where a user has a living need that has never or rarely been fulfilled on Meituan platform, he/she probably won't launch Meituan mobile app, as they may not be aware that this need can be satisfied on Meituan or may not be accustomed to fulfilling this need on Meituan. So, if our engine runs in the background of the phone and detects the user's potential living need, it will send a message pop-up to the user with a recommendation for a life service solution, if the user allows it. The message pop-up recommends the user a life service that is in the category of the highest score output by our system.

In brief, with the ability of potential needs prediction, NEON is deployed for message pop-up recommendation.
\subsubsection{Experiment Setting}
We conduct online A/B tests involving one million users over a period of two weeks. The users are randomly divided into two buckets of similar volume and are assigned different strategies for selecting items to be sent in message pop-ups.  For the first bucket, the message pop-up recommends items that belong to the category with the highest score output by our system. For the second bucket, in each hour of the day, we calculate the popularity and average CTR of each category. We distribute the traffic for message pop-ups to the categories with the most popularity and highest average CTR. 

We use CTR, CVR, CTCVR, OV, \textit{Number of Cold-start Customers} (NCC) to measure the performance of our system in potential need prediction. NCC represents the number of customers who purchase a lifestyle service that they have never acquired on the platform previously.
\subsubsection{Performance}
The results of the online A/B tests are shown in Table~\ref{tab::online3overall}. We can observe that:
\begin{itemize}[leftmargin=*]
\item{By replacing the algorithm based on popularity and average CTR with our living need prediction system NEON to determine the category of recommendation in message pop-up, all metrics show significant improvement. CTR, CVR, CTCVR, and OV increase by 8.21\%, 78.64\%, 85.71\%, and 95.92\%, respectively.} 
\item{NCC increases by 74.26\%, indicating that the online deployment of our system in message pop-up recommendations results in a 74.26\% increase in the number of customers purchasing a life service that they have not previously acquired on the Meituan platform, via the message pop-up feature.} 
\end{itemize}
Our system is able to accurately detect the potential needs of users for a specific lifestyle service, even in instances where they have never previously purchased that service on the Meituan platform, and accordingly deliver targeted message pop-ups to them. This result strongly demonstrates the exceptional capability of our system in predicting users' potential needs.
\begin{table}[]
\caption{Results of A/B tests on message pop-up recommendation.}
\vspace{-0.2cm}
\label{tab::online3overall}
\begin{tabular}{cccccc}
\hline
Metric      & CTR     & CVR      & CTCVR    & OV       & NCC      \\ \hline
Improv. & +8.21\% & +78.64\% & +85.71\% & +95.92\% & +74.26\% \\ \hline
\end{tabular}
\end{table}

In summary, the success achieved by our NEON system on three downstream recommendation tasks proves its effectiveness on fine-grained need prediction, needs-meeting way prediction, and potential need prediction, respectively. The significant performance gain in real applications can lead to huge benefits.

\section{Related Work}\label{sec::relatedwork}
As discussed above, in this work we explore the problem of living needs prediction, defined as predicting the specific living \textbf{needs} of a user given the \textbf{spatiotemporal} context. Thus, there are two closely related research topics: demand forecasting and spatiotemporal activity prediction.

\subsection{Demand Forecasting} 
Demand forecasting aims at predicting the quantity of a product or service that consumers will purchase. It helps in making informed decisions on inventory management, production scheduling, pricing strategy, etc. The problem of demand forecasting is broad and multifaceted, affecting many different industries, including restaurant~\cite{lasek2016restaurant}, manufacturing~\cite{seyedan2020predictive}, retail~\cite{ren2020demand}, tourism~\cite{song2019review}, energy~\cite{hernandez2014survey}, transportation~\cite{tskeris2011demand}, etc. 

To address the problem of demand forecasting, researchers have proposed various methods which can be broadly classified into three categories: statistical models~\cite{lasek2016restaurant,fattah2018forecasting,huber2017cluster,priyadarshi2019demand,wang2019selection}, machine learning models~\cite{reynolds2013econometric,sellers2010predicting,aishwarya2020food,ma2016demand}, and deep learning models~\cite{raza2017prediction,duncan2015probabilistic,lakshmanan2020sales,kilimci2019improved}. Statistical models, such as exponential smoothing~\cite{silva2019demand}, are well-suited for long-term demand forecasting as they are based on historical trends and patterns. However, they are not adept at handling variations or outliers in the data, making them unsuitable for volatile or short-term demand forecasting. On the other hand, machine learning models for demand forecasting, such as Random Forest based models~\cite{ramya2020advanced,abbasi2019short}, are efficient at short-term demand forecasting, but their performance drops when it comes to long-term forecasting. Deep learning models such as LSTM based models~\cite{lakshmanan2020sales} and GAN based models~\cite{husein2019generative} have the ability to capture complex patterns and dependencies in the data, making them suitable for both short-term and long-term demand forecasting. However, they require a large amount of data to work well. 

Existing demand forecasting methods can not handle the problem of living needs prediction. These methods focus on the overall demand for a particular product or service in a market, but in this work, we aim at predicting the need of a specific consumer. What's more, demand forecasting methods predict demands in several months or years, while in this work we predict a user's need at a specific time and location.

\subsection{Spatiotemporal Activity Prediction}
 Spatiotemporal activity prediction aims to predict the activity of a user at a given time and location. Previous works have employed various methods to perform spatiotemporal activity prediction. One popular approach is to build a tensor using historical data and then conduct tensor factorization to learn intrinsic association~\cite{fan2019personalized,zheng2010collaborative,bhargava2015and}. For example, Fan \textit{et al.}~\cite{fan2019personalized} propose to integrate tensor factorization with transfer learning for online activity prediction. Additionally, WDGTC~\cite{li2020tensor} proposes a low-rank tensor decomposition and completion framework for passenger flow prediction by introducing L1-norm and Graph Laplacian penalties. Recently, researchers have introduced Graph Convolutional Networks~\cite{kipf2016semi} (GCNs) to achieve high performance in spatiotemporal activity prediction. For example, SA-GCN~\cite{yu2020semantic} develops a Graph Convolutional Network with meta path-based objective function for App-usage prediction task. Furthermore, DisenHCN~\cite{li2022disenhcn} utilizes a heterogeneous hypergraph to model fine-grained user similarities, resulting in significant performance gains.

However, existing works on spatiotemporal activity prediction focus on predicting the specific activities of people, while our work focuses on the general living needs which are the driving force behind specific consumption behaviors. What's more, these works mainly focus on either online or offline activities, but in our work, we predict living needs can be satisfied both in store (offline) and via delivery (online) by different kinds of life services, which is beyond the capabilities of existing methods. 
\section{Conclusion and Future Work}\label{sec::conclusion}
In this work, we approach the new problem of living needs prediction, which is critical in life services platforms.
We present the NEON system in Meituan, consisting of three phases, feature mining, feature fusion, and multitask prediction. 
Large-scale online A/B testing in three downstream applications, along with extensive offline evaluation, strongly confirm the effectiveness of our system.
As for future work, we plan to test NEON's performance in more downstream applications.

\section*{Acknowledgement}
This work is supported in part by National Key Research and Development Program of China under 2022YFB3104702. This work is supported in part by National Natural Science Foundation of China under 62272262, 61971267, and 61972223. This work is supported in part by a grant from the Guoqiang Institute, Tsinghua University under 2021GQG1005. This work is supported in part by Beijing National Research Center for Information Science and Technology. This work is also supported by Meituan.
\clearpage
\balance

\bibliographystyle{ACM-Reference-Format}
\bibliography{bibliography}


\begin{thebibliography}{42}


\ifx \showCODEN    \undefined \def \showCODEN     #1{\unskip}     \fi
\ifx \showDOI      \undefined \def \showDOI       #1{#1}\fi
\ifx \showISBNx    \undefined \def \showISBNx     #1{\unskip}     \fi
\ifx \showISBNxiii \undefined \def \showISBNxiii  #1{\unskip}     \fi
\ifx \showISSN     \undefined \def \showISSN      #1{\unskip}     \fi
\ifx \showLCCN     \undefined \def \showLCCN      #1{\unskip}     \fi
\ifx \shownote     \undefined \def \shownote      #1{#1}          \fi
\ifx \showarticletitle \undefined \def \showarticletitle #1{#1}   \fi
\ifx \showURL      \undefined \def \showURL       {\relax}        \fi
\providecommand\bibfield[2]{#2}
\providecommand\bibinfo[2]{#2}
\providecommand\natexlab[1]{#1}
\providecommand\showeprint[2][]{arXiv:#2}

\bibitem[\protect\citeauthoryear{Abbasi, Javaid, Ghuman, Khan, and
  Ur~Rehman}{Abbasi et~al\mbox{.}}{2019}]%
        {abbasi2019short}
\bibfield{author}{\bibinfo{person}{Raza~Abid Abbasi}, \bibinfo{person}{Nadeem
  Javaid}, \bibinfo{person}{Muhammad Nauman~Javid Ghuman},
  \bibinfo{person}{Zahoor~Ali Khan}, {and} \bibinfo{person}{Shujat Ur~Rehman}.}
  \bibinfo{year}{2019}\natexlab{}.
\newblock \showarticletitle{Short term load forecasting using XGBoost}. In
  \bibinfo{booktitle}{{\em Web, Artificial Intelligence and Network
  Applications: Proceedings of the Workshops of the 33rd International
  Conference on Advanced Information Networking and Applications (WAINA-2019)
  33}}. Springer, \bibinfo{pages}{1120--1131}.
\newblock


\bibitem[\protect\citeauthoryear{Aishwarya, Rao, Kumari, Mishra, and
  Rashmi}{Aishwarya et~al\mbox{.}}{2020}]%
        {aishwarya2020food}
\bibfield{author}{\bibinfo{person}{K Aishwarya}, \bibinfo{person}{N Rao},
  \bibinfo{person}{Nikita Kumari}, \bibinfo{person}{Akshit Mishra}, {and}
  \bibinfo{person}{MR Rashmi}.} \bibinfo{year}{2020}\natexlab{}.
\newblock \showarticletitle{Food Demand Prediction using machine learning}.
\newblock \bibinfo{journal}{{\em International Research Journal of Engineering
  and Technology (IRJET)\/}}  \bibinfo{volume}{7} (\bibinfo{year}{2020}).
\newblock


\bibitem[\protect\citeauthoryear{Bhargava, Phan, Zhou, and Lee}{Bhargava
  et~al\mbox{.}}{2015}]%
        {bhargava2015and}
\bibfield{author}{\bibinfo{person}{Preeti Bhargava}, \bibinfo{person}{Thomas
  Phan}, \bibinfo{person}{Jiayu Zhou}, {and} \bibinfo{person}{Juhan Lee}.}
  \bibinfo{year}{2015}\natexlab{}.
\newblock \showarticletitle{Who, what, when, and where: Multi-dimensional
  collaborative recommendations using tensor factorization on sparse
  user-generated data}. In \bibinfo{booktitle}{{\em WWW}}.
  \bibinfo{pages}{130--140}.
\newblock


\bibitem[\protect\citeauthoryear{Cheng, Yang, King, and Lyu}{Cheng
  et~al\mbox{.}}{2012}]%
        {cheng2012fused}
\bibfield{author}{\bibinfo{person}{Chen Cheng}, \bibinfo{person}{Haiqin Yang},
  \bibinfo{person}{Irwin King}, {and} \bibinfo{person}{Michael Lyu}.}
  \bibinfo{year}{2012}\natexlab{}.
\newblock \showarticletitle{Fused matrix factorization with geographical and
  social influence in location-based social networks}. In
  \bibinfo{booktitle}{{\em Proceedings of the AAAI conference on artificial
  intelligence}}, Vol.~\bibinfo{volume}{26}. \bibinfo{pages}{17--23}.
\newblock


\bibitem[\protect\citeauthoryear{Cheng, Koc, Harmsen, Shaked, Chandra, Aradhye,
  Anderson, Corrado, Chai, Ispir, et~al\mbox{.}}{Cheng et~al\mbox{.}}{2016}]%
        {cheng2016wide}
\bibfield{author}{\bibinfo{person}{Heng-Tze Cheng}, \bibinfo{person}{Levent
  Koc}, \bibinfo{person}{Jeremiah Harmsen}, \bibinfo{person}{Tal Shaked},
  \bibinfo{person}{Tushar Chandra}, \bibinfo{person}{Hrishi Aradhye},
  \bibinfo{person}{Glen Anderson}, \bibinfo{person}{Greg Corrado},
  \bibinfo{person}{Wei Chai}, \bibinfo{person}{Mustafa Ispir}, {et~al\mbox{.}}}
  \bibinfo{year}{2016}\natexlab{}.
\newblock \showarticletitle{Wide \& deep learning for recommender systems}. In
  \bibinfo{booktitle}{{\em Proceedings of the 1st workshop on deep learning for
  recommender systems}}. \bibinfo{pages}{7--10}.
\newblock


\bibitem[\protect\citeauthoryear{Duncan and Elkan}{Duncan and Elkan}{2015}]%
        {duncan2015probabilistic}
\bibfield{author}{\bibinfo{person}{Brendan~Andrew Duncan} {and}
  \bibinfo{person}{Charles~Peter Elkan}.} \bibinfo{year}{2015}\natexlab{}.
\newblock \showarticletitle{Probabilistic modeling of a sales funnel to
  prioritize leads}. In \bibinfo{booktitle}{{\em Proceedings of the 21th ACM
  SIGKDD international conference on knowledge discovery and data mining}}.
  \bibinfo{pages}{1751--1758}.
\newblock


\bibitem[\protect\citeauthoryear{Fan, Tu, Li, Chen, Gao, Zhang, Su, and
  Jin}{Fan et~al\mbox{.}}{2019}]%
        {fan2019personalized}
\bibfield{author}{\bibinfo{person}{Yali Fan}, \bibinfo{person}{Zhen Tu},
  \bibinfo{person}{Yong Li}, \bibinfo{person}{Xiang Chen}, \bibinfo{person}{Hui
  Gao}, \bibinfo{person}{Lin Zhang}, \bibinfo{person}{Li Su}, {and}
  \bibinfo{person}{Depeng Jin}.} \bibinfo{year}{2019}\natexlab{}.
\newblock \showarticletitle{Personalized context-aware collaborative online
  activity prediction}.
\newblock \bibinfo{journal}{{\em UbiComp\/}} \bibinfo{volume}{3},
  \bibinfo{number}{4} (\bibinfo{year}{2019}), \bibinfo{pages}{1--28}.
\newblock


\bibitem[\protect\citeauthoryear{Fattah, Ezzine, Aman, El~Moussami, and
  Lachhab}{Fattah et~al\mbox{.}}{2018}]%
        {fattah2018forecasting}
\bibfield{author}{\bibinfo{person}{Jamal Fattah}, \bibinfo{person}{Latifa
  Ezzine}, \bibinfo{person}{Zineb Aman}, \bibinfo{person}{Haj El~Moussami},
  {and} \bibinfo{person}{Abdeslam Lachhab}.} \bibinfo{year}{2018}\natexlab{}.
\newblock \showarticletitle{Forecasting of demand using ARIMA model}.
\newblock \bibinfo{journal}{{\em International Journal of Engineering Business
  Management\/}}  \bibinfo{volume}{10} (\bibinfo{year}{2018}),
  \bibinfo{pages}{1847979018808673}.
\newblock


\bibitem[\protect\citeauthoryear{Glorot, Bordes, and Bengio}{Glorot
  et~al\mbox{.}}{2011}]%
        {glorot2011deep}
\bibfield{author}{\bibinfo{person}{Xavier Glorot}, \bibinfo{person}{Antoine
  Bordes}, {and} \bibinfo{person}{Yoshua Bengio}.}
  \bibinfo{year}{2011}\natexlab{}.
\newblock \showarticletitle{Deep sparse rectifier neural networks}. In
  \bibinfo{booktitle}{{\em Proceedings of the fourteenth international
  conference on artificial intelligence and statistics}}. JMLR Workshop and
  Conference Proceedings, \bibinfo{pages}{315--323}.
\newblock


\bibitem[\protect\citeauthoryear{Hernandez, Baladron, Aguiar, Carro,
  Sanchez-Esguevillas, Lloret, and Massana}{Hernandez et~al\mbox{.}}{2014}]%
        {hernandez2014survey}
\bibfield{author}{\bibinfo{person}{Luis Hernandez}, \bibinfo{person}{Carlos
  Baladron}, \bibinfo{person}{Javier~M Aguiar}, \bibinfo{person}{Bel{\'e}n
  Carro}, \bibinfo{person}{Antonio~J Sanchez-Esguevillas},
  \bibinfo{person}{Jaime Lloret}, {and} \bibinfo{person}{Joaquim Massana}.}
  \bibinfo{year}{2014}\natexlab{}.
\newblock \showarticletitle{A survey on electric power demand forecasting:
  future trends in smart grids, microgrids and smart buildings}.
\newblock \bibinfo{journal}{{\em IEEE Communications Surveys \& Tutorials\/}}
  \bibinfo{volume}{16}, \bibinfo{number}{3} (\bibinfo{year}{2014}),
  \bibinfo{pages}{1460--1495}.
\newblock


\bibitem[\protect\citeauthoryear{Huber, Gossmann, and Stuckenschmidt}{Huber
  et~al\mbox{.}}{2017}]%
        {huber2017cluster}
\bibfield{author}{\bibinfo{person}{Jakob Huber}, \bibinfo{person}{Alexander
  Gossmann}, {and} \bibinfo{person}{Heiner Stuckenschmidt}.}
  \bibinfo{year}{2017}\natexlab{}.
\newblock \showarticletitle{Cluster-based hierarchical demand forecasting for
  perishable goods}.
\newblock \bibinfo{journal}{{\em Expert systems with applications\/}}
  \bibinfo{volume}{76} (\bibinfo{year}{2017}), \bibinfo{pages}{140--151}.
\newblock


\bibitem[\protect\citeauthoryear{Husein, Arsyal, Sinaga, and Syahputa}{Husein
  et~al\mbox{.}}{2019}]%
        {husein2019generative}
\bibfield{author}{\bibinfo{person}{Amir~Mahmud Husein},
  \bibinfo{person}{Muhammad Arsyal}, \bibinfo{person}{Sutrisno Sinaga}, {and}
  \bibinfo{person}{Hendra Syahputa}.} \bibinfo{year}{2019}\natexlab{}.
\newblock \showarticletitle{Generative adversarial networks time series models
  to forecast medicine daily sales in hospital}.
\newblock \bibinfo{journal}{{\em Sinkron: jurnal dan penelitian teknik
  informatika\/}} \bibinfo{volume}{3}, \bibinfo{number}{2}
  (\bibinfo{year}{2019}), \bibinfo{pages}{112--118}.
\newblock


\bibitem[\protect\citeauthoryear{Ioffe and Szegedy}{Ioffe and Szegedy}{2015}]%
        {ioffe2015batch}
\bibfield{author}{\bibinfo{person}{Sergey Ioffe} {and}
  \bibinfo{person}{Christian Szegedy}.} \bibinfo{year}{2015}\natexlab{}.
\newblock \showarticletitle{Batch normalization: Accelerating deep network
  training by reducing internal covariate shift}. In \bibinfo{booktitle}{{\em
  International conference on machine learning}}. pmlr,
  \bibinfo{pages}{448--456}.
\newblock


\bibitem[\protect\citeauthoryear{Kilimci, Akyuz, Uysal, Akyokus, Uysal,
  Atak~Bulbul, and Ekmis}{Kilimci et~al\mbox{.}}{2019}]%
        {kilimci2019improved}
\bibfield{author}{\bibinfo{person}{Zeynep~Hilal Kilimci},
  \bibinfo{person}{A~Okay Akyuz}, \bibinfo{person}{Mitat Uysal},
  \bibinfo{person}{Selim Akyokus}, \bibinfo{person}{M~Ozan Uysal},
  \bibinfo{person}{Berna Atak~Bulbul}, {and} \bibinfo{person}{Mehmet~Ali
  Ekmis}.} \bibinfo{year}{2019}\natexlab{}.
\newblock \showarticletitle{An improved demand forecasting model using deep
  learning approach and proposed decision integration strategy for supply
  chain}.
\newblock \bibinfo{journal}{{\em Complexity\/}}  \bibinfo{volume}{2019}
  (\bibinfo{year}{2019}).
\newblock


\bibitem[\protect\citeauthoryear{Kipf and Welling}{Kipf and Welling}{2016}]%
        {kipf2016semi}
\bibfield{author}{\bibinfo{person}{Thomas~N Kipf} {and} \bibinfo{person}{Max
  Welling}.} \bibinfo{year}{2016}\natexlab{}.
\newblock \showarticletitle{Semi-supervised classification with graph
  convolutional networks}.
\newblock \bibinfo{journal}{{\em arXiv preprint arXiv:1609.02907\/}}
  (\bibinfo{year}{2016}).
\newblock


\bibitem[\protect\citeauthoryear{Kullback and Leibler}{Kullback and
  Leibler}{1951}]%
        {kullback1951information}
\bibfield{author}{\bibinfo{person}{Solomon Kullback} {and}
  \bibinfo{person}{Richard~A Leibler}.} \bibinfo{year}{1951}\natexlab{}.
\newblock \showarticletitle{On information and sufficiency}.
\newblock \bibinfo{journal}{{\em The annals of mathematical statistics\/}}
  \bibinfo{volume}{22}, \bibinfo{number}{1} (\bibinfo{year}{1951}),
  \bibinfo{pages}{79--86}.
\newblock


\bibitem[\protect\citeauthoryear{Lakshmanan, Vivek~Raja, and
  Kalathiappan}{Lakshmanan et~al\mbox{.}}{2020}]%
        {lakshmanan2020sales}
\bibfield{author}{\bibinfo{person}{Balakrishnan Lakshmanan},
  \bibinfo{person}{Palaniappan Senthil~Nayagam Vivek~Raja}, {and}
  \bibinfo{person}{Viswanathan Kalathiappan}.} \bibinfo{year}{2020}\natexlab{}.
\newblock \showarticletitle{Sales demand forecasting using LSTM network}. In
  \bibinfo{booktitle}{{\em Artificial Intelligence and Evolutionary
  Computations in Engineering Systems}}. Springer, \bibinfo{pages}{125--132}.
\newblock


\bibitem[\protect\citeauthoryear{Lasek, Cercone, and Saunders}{Lasek
  et~al\mbox{.}}{2016}]%
        {lasek2016restaurant}
\bibfield{author}{\bibinfo{person}{Agnieszka Lasek}, \bibinfo{person}{Nick
  Cercone}, {and} \bibinfo{person}{Jim Saunders}.}
  \bibinfo{year}{2016}\natexlab{}.
\newblock \showarticletitle{Restaurant sales and customer demand forecasting:
  Literature survey and categorization of methods}.
\newblock \bibinfo{journal}{{\em Smart City 360°: First EAI International
  Summit, Smart City 360°, Bratislava, Slovakia and Toronto, Canada, October
  13-16, 2015. Revised Selected Papers 1\/}} (\bibinfo{year}{2016}),
  \bibinfo{pages}{479--491}.
\newblock


\bibitem[\protect\citeauthoryear{Li, Gao, Yao, Li, Jin, and Li}{Li
  et~al\mbox{.}}{2022}]%
        {li2022disenhcn}
\bibfield{author}{\bibinfo{person}{Yinfeng Li}, \bibinfo{person}{Chen Gao},
  \bibinfo{person}{Quanming Yao}, \bibinfo{person}{Tong Li},
  \bibinfo{person}{Depeng Jin}, {and} \bibinfo{person}{Yong Li}.}
  \bibinfo{year}{2022}\natexlab{}.
\newblock \showarticletitle{DisenHCN: Disentangled Hypergraph Convolutional
  Networks for Spatiotemporal Activity Prediction}.
\newblock \bibinfo{journal}{{\em arXiv preprint arXiv:2208.06794\/}}
  (\bibinfo{year}{2022}).
\newblock


\bibitem[\protect\citeauthoryear{Li, Sergin, Yan, Zhang, and Tsung}{Li
  et~al\mbox{.}}{2020}]%
        {li2020tensor}
\bibfield{author}{\bibinfo{person}{Ziyue Li},
  \bibinfo{person}{Nurettin~Dorukhan Sergin}, \bibinfo{person}{Hao Yan},
  \bibinfo{person}{Chen Zhang}, {and} \bibinfo{person}{Fugee Tsung}.}
  \bibinfo{year}{2020}\natexlab{}.
\newblock \showarticletitle{Tensor completion for weakly-dependent data on
  graph for metro passenger flow prediction}. In \bibinfo{booktitle}{{\em
  proceedings of the AAAI conference on artificial intelligence}},
  Vol.~\bibinfo{volume}{34}. \bibinfo{pages}{4804--4810}.
\newblock


\bibitem[\protect\citeauthoryear{Liu, Zhu, Li, Zhang, Lai, Tang, He, Li, and
  Yu}{Liu et~al\mbox{.}}{2020}]%
        {liu2020autofis}
\bibfield{author}{\bibinfo{person}{Bin Liu}, \bibinfo{person}{Chenxu Zhu},
  \bibinfo{person}{Guilin Li}, \bibinfo{person}{Weinan Zhang},
  \bibinfo{person}{Jincai Lai}, \bibinfo{person}{Ruiming Tang},
  \bibinfo{person}{Xiuqiang He}, \bibinfo{person}{Zhenguo Li}, {and}
  \bibinfo{person}{Yong Yu}.} \bibinfo{year}{2020}\natexlab{}.
\newblock \showarticletitle{Autofis: Automatic feature interaction selection in
  factorization models for click-through rate prediction}. In
  \bibinfo{booktitle}{{\em Proceedings of the 26th ACM SIGKDD International
  Conference on Knowledge Discovery \& Data Mining}}.
  \bibinfo{pages}{2636--2645}.
\newblock


\bibitem[\protect\citeauthoryear{Ma, Zhao, Yi, Chen, Hong, and Chi}{Ma
  et~al\mbox{.}}{2018b}]%
        {ma2018modeling}
\bibfield{author}{\bibinfo{person}{Jiaqi Ma}, \bibinfo{person}{Zhe Zhao},
  \bibinfo{person}{Xinyang Yi}, \bibinfo{person}{Jilin Chen},
  \bibinfo{person}{Lichan Hong}, {and} \bibinfo{person}{Ed~H Chi}.}
  \bibinfo{year}{2018}\natexlab{b}.
\newblock \showarticletitle{Modeling task relationships in multi-task learning
  with multi-gate mixture-of-experts}. In \bibinfo{booktitle}{{\em Proceedings
  of the 24th ACM SIGKDD international conference on knowledge discovery \&
  data mining}}. \bibinfo{pages}{1930--1939}.
\newblock


\bibitem[\protect\citeauthoryear{Ma, Fildes, and Huang}{Ma
  et~al\mbox{.}}{2016}]%
        {ma2016demand}
\bibfield{author}{\bibinfo{person}{Shaohui Ma}, \bibinfo{person}{Robert
  Fildes}, {and} \bibinfo{person}{Tao Huang}.} \bibinfo{year}{2016}\natexlab{}.
\newblock \showarticletitle{Demand forecasting with high dimensional data: The
  case of SKU retail sales forecasting with intra-and inter-category
  promotional information}.
\newblock \bibinfo{journal}{{\em European Journal of Operational Research\/}}
  \bibinfo{volume}{249}, \bibinfo{number}{1} (\bibinfo{year}{2016}),
  \bibinfo{pages}{245--257}.
\newblock


\bibitem[\protect\citeauthoryear{Ma, Zhao, Huang, Wang, Hu, Zhu, and Gai}{Ma
  et~al\mbox{.}}{2018a}]%
        {ma2018entire}
\bibfield{author}{\bibinfo{person}{Xiao Ma}, \bibinfo{person}{Liqin Zhao},
  \bibinfo{person}{Guan Huang}, \bibinfo{person}{Zhi Wang},
  \bibinfo{person}{Zelin Hu}, \bibinfo{person}{Xiaoqiang Zhu}, {and}
  \bibinfo{person}{Kun Gai}.} \bibinfo{year}{2018}\natexlab{a}.
\newblock \showarticletitle{Entire space multi-task model: An effective
  approach for estimating post-click conversion rate}. In
  \bibinfo{booktitle}{{\em The 41st International ACM SIGIR Conference on
  Research \& Development in Information Retrieval}}.
  \bibinfo{pages}{1137--1140}.
\newblock


\bibitem[\protect\citeauthoryear{Palumbo, Rizzo, Troncy, Baralis, Osella, and
  Ferro}{Palumbo et~al\mbox{.}}{2018}]%
        {palumbo2018knowledge}
\bibfield{author}{\bibinfo{person}{Enrico Palumbo}, \bibinfo{person}{Giuseppe
  Rizzo}, \bibinfo{person}{Rapha{\"e}l Troncy}, \bibinfo{person}{Elena
  Baralis}, \bibinfo{person}{Michele Osella}, {and} \bibinfo{person}{Enrico
  Ferro}.} \bibinfo{year}{2018}\natexlab{}.
\newblock \showarticletitle{Knowledge graph embeddings with node2vec for item
  recommendation}. In \bibinfo{booktitle}{{\em The Semantic Web: ESWC 2018
  Satellite Events: ESWC 2018 Satellite Events, Heraklion, Crete, Greece, June
  3-7, 2018, Revised Selected Papers 15}}. Springer, \bibinfo{pages}{117--120}.
\newblock


\bibitem[\protect\citeauthoryear{Priyadarshi, Panigrahi, Routroy, and
  Garg}{Priyadarshi et~al\mbox{.}}{2019}]%
        {priyadarshi2019demand}
\bibfield{author}{\bibinfo{person}{Rahul Priyadarshi}, \bibinfo{person}{Akash
  Panigrahi}, \bibinfo{person}{Srikanta Routroy}, {and}
  \bibinfo{person}{Girish~Kant Garg}.} \bibinfo{year}{2019}\natexlab{}.
\newblock \showarticletitle{Demand forecasting at retail stage for selected
  vegetables: a performance analysis}.
\newblock \bibinfo{journal}{{\em Journal of Modelling in Management\/}}
  \bibinfo{volume}{14}, \bibinfo{number}{4} (\bibinfo{year}{2019}),
  \bibinfo{pages}{1042--1063}.
\newblock


\bibitem[\protect\citeauthoryear{Ramya and Vedavathi}{Ramya and
  Vedavathi}{2020}]%
        {ramya2020advanced}
\bibfield{author}{\bibinfo{person}{B~Sri~Sai Ramya} {and} \bibinfo{person}{K
  Vedavathi}.} \bibinfo{year}{2020}\natexlab{}.
\newblock \showarticletitle{An advanced sales forecasting using machine
  learning algorithm}.
\newblock \bibinfo{journal}{{\em International Journal of Innovative Science
  and Research Technology\/}} \bibinfo{volume}{5}, \bibinfo{number}{5}
  (\bibinfo{year}{2020}), \bibinfo{pages}{342--345}.
\newblock


\bibitem[\protect\citeauthoryear{Raza}{Raza}{2017}]%
        {raza2017prediction}
\bibfield{author}{\bibinfo{person}{Kamran Raza}.}
  \bibinfo{year}{2017}\natexlab{}.
\newblock \showarticletitle{Prediction of Stock Market performance by using
  machine learning techniques}. In \bibinfo{booktitle}{{\em 2017 International
  conference on innovations in electrical engineering and computational
  technologies (ICIEECT)}}. IEEE, \bibinfo{pages}{1--1}.
\newblock


\bibitem[\protect\citeauthoryear{Ren, Chan, and Siqin}{Ren
  et~al\mbox{.}}{2020}]%
        {ren2020demand}
\bibfield{author}{\bibinfo{person}{Shuyun Ren}, \bibinfo{person}{Hau-Ling
  Chan}, {and} \bibinfo{person}{Tana Siqin}.} \bibinfo{year}{2020}\natexlab{}.
\newblock \showarticletitle{Demand forecasting in retail operations for
  fashionable products: methods, practices, and real case study}.
\newblock \bibinfo{journal}{{\em Annals of Operations Research\/}}
  \bibinfo{volume}{291} (\bibinfo{year}{2020}), \bibinfo{pages}{761--777}.
\newblock


\bibitem[\protect\citeauthoryear{Reynolds, Rahman, and Balinbin}{Reynolds
  et~al\mbox{.}}{2013}]%
        {reynolds2013econometric}
\bibfield{author}{\bibinfo{person}{Dennis Reynolds}, \bibinfo{person}{Imran
  Rahman}, {and} \bibinfo{person}{William Balinbin}.}
  \bibinfo{year}{2013}\natexlab{}.
\newblock \showarticletitle{Econometric modeling of the US restaurant
  industry}.
\newblock \bibinfo{journal}{{\em International Journal of Hospitality
  Management\/}}  \bibinfo{volume}{34} (\bibinfo{year}{2013}),
  \bibinfo{pages}{317--323}.
\newblock


\bibitem[\protect\citeauthoryear{Sellers and Shmueli}{Sellers and
  Shmueli}{2010}]%
        {sellers2010predicting}
\bibfield{author}{\bibinfo{person}{Kimberly~F Sellers} {and}
  \bibinfo{person}{Galit Shmueli}.} \bibinfo{year}{2010}\natexlab{}.
\newblock \showarticletitle{Predicting censored count data with COM-Poisson
  regression}.
\newblock \bibinfo{journal}{{\em Robert H. Smith School Research Paper No.
  RHS-06-129\/}} (\bibinfo{year}{2010}).
\newblock


\bibitem[\protect\citeauthoryear{Seyedan and Mafakheri}{Seyedan and
  Mafakheri}{2020}]%
        {seyedan2020predictive}
\bibfield{author}{\bibinfo{person}{Mahya Seyedan} {and}
  \bibinfo{person}{Fereshteh Mafakheri}.} \bibinfo{year}{2020}\natexlab{}.
\newblock \showarticletitle{Predictive big data analytics for supply chain
  demand forecasting: methods, applications, and research opportunities}.
\newblock \bibinfo{journal}{{\em Journal of Big Data\/}} \bibinfo{volume}{7},
  \bibinfo{number}{1} (\bibinfo{year}{2020}), \bibinfo{pages}{1--22}.
\newblock


\bibitem[\protect\citeauthoryear{Silva, Figueiredo, and Braga}{Silva
  et~al\mbox{.}}{2019}]%
        {silva2019demand}
\bibfield{author}{\bibinfo{person}{Juliana~C Silva}, \bibinfo{person}{Manuel~C
  Figueiredo}, {and} \bibinfo{person}{Ana~C Braga}.}
  \bibinfo{year}{2019}\natexlab{}.
\newblock \showarticletitle{Demand forecasting: A case study in the food
  industry}. In \bibinfo{booktitle}{{\em Computational Science and Its
  Applications--ICCSA 2019: 19th International Conference, Saint Petersburg,
  Russia, July 1--4, 2019, Proceedings, Part III 19}}. Springer,
  \bibinfo{pages}{50--63}.
\newblock


\bibitem[\protect\citeauthoryear{Song, Qiu, and Park}{Song
  et~al\mbox{.}}{2019}]%
        {song2019review}
\bibfield{author}{\bibinfo{person}{Haiyan Song}, \bibinfo{person}{Richard~TR
  Qiu}, {and} \bibinfo{person}{Jinah Park}.} \bibinfo{year}{2019}\natexlab{}.
\newblock \showarticletitle{A review of research on tourism demand forecasting:
  Launching the Annals of Tourism Research Curated Collection on tourism demand
  forecasting}.
\newblock \bibinfo{journal}{{\em Annals of Tourism Research\/}}
  \bibinfo{volume}{75} (\bibinfo{year}{2019}), \bibinfo{pages}{338--362}.
\newblock


\bibitem[\protect\citeauthoryear{Tang, Liu, Zhao, and Gong}{Tang
  et~al\mbox{.}}{2020}]%
        {tang2020progressive}
\bibfield{author}{\bibinfo{person}{Hongyan Tang}, \bibinfo{person}{Junning
  Liu}, \bibinfo{person}{Ming Zhao}, {and} \bibinfo{person}{Xudong Gong}.}
  \bibinfo{year}{2020}\natexlab{}.
\newblock \showarticletitle{Progressive layered extraction (ple): A novel
  multi-task learning (mtl) model for personalized recommendations}. In
  \bibinfo{booktitle}{{\em Proceedings of the 14th ACM Conference on
  Recommender Systems}}. \bibinfo{pages}{269--278}.
\newblock


\bibitem[\protect\citeauthoryear{Tskeris and Tsekeris}{Tskeris and
  Tsekeris}{2011}]%
        {tskeris2011demand}
\bibfield{author}{\bibinfo{person}{Theodore Tskeris} {and}
  \bibinfo{person}{Charalambos Tsekeris}.} \bibinfo{year}{2011}\natexlab{}.
\newblock \showarticletitle{Demand forecasting in transport: Overview and
  modeling advances}.
\newblock \bibinfo{journal}{{\em Economic research-Ekonomska
  istra{\v{z}}ivanja\/}} \bibinfo{volume}{24}, \bibinfo{number}{1}
  (\bibinfo{year}{2011}), \bibinfo{pages}{82--94}.
\newblock


\bibitem[\protect\citeauthoryear{Wang, Liu, and Liu}{Wang
  et~al\mbox{.}}{2019}]%
        {wang2019selection}
\bibfield{author}{\bibinfo{person}{Jiaxing Wang}, \bibinfo{person}{GQ Liu},
  {and} \bibinfo{person}{Lu Liu}.} \bibinfo{year}{2019}\natexlab{}.
\newblock \showarticletitle{A selection of advanced technologies for demand
  forecasting in the retail industry}. In \bibinfo{booktitle}{{\em 2019 IEEE
  4th International Conference on Big Data Analytics (ICBDA)}}. IEEE,
  \bibinfo{pages}{317--320}.
\newblock


\bibitem[\protect\citeauthoryear{Wang, Fu, Fu, and Wang}{Wang
  et~al\mbox{.}}{2017}]%
        {wang2017deep}
\bibfield{author}{\bibinfo{person}{Ruoxi Wang}, \bibinfo{person}{Bin Fu},
  \bibinfo{person}{Gang Fu}, {and} \bibinfo{person}{Mingliang Wang}.}
  \bibinfo{year}{2017}\natexlab{}.
\newblock \showarticletitle{Deep \& cross network for ad click predictions}.
\newblock In \bibinfo{booktitle}{{\em Proceedings of the ADKDD'17}}.
  \bibinfo{pages}{1--7}.
\newblock


\bibitem[\protect\citeauthoryear{Wen, Zhang, Wang, Lv, Bao, Lin, and Yang}{Wen
  et~al\mbox{.}}{2020}]%
        {wen2020entire}
\bibfield{author}{\bibinfo{person}{Hong Wen}, \bibinfo{person}{Jing Zhang},
  \bibinfo{person}{Yuan Wang}, \bibinfo{person}{Fuyu Lv},
  \bibinfo{person}{Wentian Bao}, \bibinfo{person}{Quan Lin}, {and}
  \bibinfo{person}{Keping Yang}.} \bibinfo{year}{2020}\natexlab{}.
\newblock \showarticletitle{Entire space multi-task modeling via post-click
  behavior decomposition for conversion rate prediction}. In
  \bibinfo{booktitle}{{\em Proceedings of the 43rd International ACM SIGIR
  conference on research and development in Information Retrieval}}.
  \bibinfo{pages}{2377--2386}.
\newblock


\bibitem[\protect\citeauthoryear{Yu, Xia, Wang, Feng, and Li}{Yu
  et~al\mbox{.}}{2020}]%
        {yu2020semantic}
\bibfield{author}{\bibinfo{person}{Yue Yu}, \bibinfo{person}{Tong Xia},
  \bibinfo{person}{Huandong Wang}, \bibinfo{person}{Jie Feng}, {and}
  \bibinfo{person}{Yong Li}.} \bibinfo{year}{2020}\natexlab{}.
\newblock \showarticletitle{Semantic-aware spatio-temporal app usage
  representation via graph convolutional network}.
\newblock \bibinfo{journal}{{\em UbiComp\/}} \bibinfo{volume}{4},
  \bibinfo{number}{3} (\bibinfo{year}{2020}), \bibinfo{pages}{1--24}.
\newblock


\bibitem[\protect\citeauthoryear{Zheng, Cao, Zheng, Xie, and Yang}{Zheng
  et~al\mbox{.}}{2010}]%
        {zheng2010collaborative}
\bibfield{author}{\bibinfo{person}{Vincent~W Zheng}, \bibinfo{person}{Bin Cao},
  \bibinfo{person}{Yu Zheng}, \bibinfo{person}{Xing Xie}, {and}
  \bibinfo{person}{Qiang Yang}.} \bibinfo{year}{2010}\natexlab{}.
\newblock \showarticletitle{Collaborative filtering meets mobile
  recommendation: A user-centered approach}. In \bibinfo{booktitle}{{\em
  Twenty-fourth AAAI conference on artificial intelligence}}.
\newblock


\bibitem[\protect\citeauthoryear{Zhou, Zhu, Song, Fan, Zhu, Ma, Yan, Jin, Li,
  and Gai}{Zhou et~al\mbox{.}}{2018}]%
        {zhou2018deep}
\bibfield{author}{\bibinfo{person}{Guorui Zhou}, \bibinfo{person}{Xiaoqiang
  Zhu}, \bibinfo{person}{Chenru Song}, \bibinfo{person}{Ying Fan},
  \bibinfo{person}{Han Zhu}, \bibinfo{person}{Xiao Ma},
  \bibinfo{person}{Yanghui Yan}, \bibinfo{person}{Junqi Jin},
  \bibinfo{person}{Han Li}, {and} \bibinfo{person}{Kun Gai}.}
  \bibinfo{year}{2018}\natexlab{}.
\newblock \showarticletitle{Deep interest network for click-through rate
  prediction}. In \bibinfo{booktitle}{{\em Proceedings of the 24th ACM SIGKDD
  international conference on knowledge discovery \& data mining}}.
  \bibinfo{pages}{1059--1068}.
\newblock


\end{thebibliography}
\balance
\clearpage

\appendix
\section{APPENDIX FOR REPRODUCIBILITY}
\subsection{Model Implementation Details}\label{sec::implementation}
In this section we provide implementation details of our system. Due to the massive amount of training data and the high demand for low latency in our online system, we simplify each network within the framework. In our final implementation, feature merging network $h^M$ is a one-layer fully-connected network of 120 output units (120D FC). User preference network $h^U$ is a 340D FC. Shared network $E^S$ and expert networks $E^W$/$E^N$ are 256D FCs. Need prediction network $t^N$ is a 10D FC. In-store/delivery classification network $t^W$ is a two-layer perceptron, which has 10 hidden units and 2 output units. Complexifying these neural networks can help us further optimize performance, but at the same time, it would lead to an increase in system latency. The activation function of all mentioned networks is ReLU~\cite{glorot2011deep}. We adopt batch normalization~\cite{ioffe2015batch} right after $h^M$, $h^U$, $E^S$, $E^P$, and $E^N$. Following such implementation, we conduct rich online and offline experiments to prove the effectiveness of our model, which are shown in Section~\ref{sec::experiments} and Section~\ref{sec::online}.
\subsection{Dataset}\label{sec::dataset}
We conduct our offline experiment on a real-world dataset at the scale of billions. The dataset is a sampling of all purchase records in 2022 on the platform. We sample the records according to the percentage of purchases of each kind of life service. The dataset includes more than 7 billion real purchase records from 65 million users.
Each instance in the dataset includes user profile, time, location, and other real-time environmental factors, and the kind of life service the user purchase. The type of life service consumed by the user reflects their actual living need in the spatiotemporal context. Following existing works~\cite{cheng2012fused,palumbo2018knowledge}, we randomly sample 80\% of the dataset as the training set, and 20\% as the test set.
\subsection{Metrics}\label{sec::metrics}
We design a metric named Sort Accuracy (SA) to measure the performance of systems on our problem. The metric SA can be defined as follows.
For user scene $i$, our system outputs scores for all kinds of living needs. We sort the living needs by their scores and get a list. We define \textit{Relative Ranking Error} as the difference between the actual ranking position and the ideal ranking position (the first position) of the ground truth need, and further define \textit{Maximum Ranking Error} as the maximum relative ranking error any system can give for a user scene (the number of categories of user needs - 1). For example, in our system which handles 10 types of living need, for user scene $i$, if the ground truth need is ranked third, then the relative ranking error is 2 (2 = 3-1), and the maximum ranking error is 9 (9 = 10-1). Then Sort Accuracy (SA) can be defined as follows,
\begin{equation}
\text{SA}=\text{average}_{i\in T}\left(1-\frac{\text{Relative Ranking Error}}{\text{Maximum Ranking Error}}\right)
\end{equation}
where $T$ denotes the testing set on which the metric is calculated. We first calculate $1-\text{Relative Ranking Error}/\text{Maximum Ranking Error}$ of every user scene in the testing set and then compute the average. We then define Via-delivery Sort Accuracy (VDSA) and In-store Sort Accuracy (ISSA) to measure systems' performance on delivery and in-store living needs.
\begin{equation}
\text{VDSA}=\text{average}_{i\in T_{VD}}\left(1-\frac{\text{Relative Ranking Error}}{\text{Maximum Ranking Error}}\right)
\end{equation}
\begin{equation}
\text{ISSA}=\text{average}_{i\in T_{IS}}\left(1-\frac{\text{Relative Ranking Error}}{\text{Maximum Ranking Error}}\right)
\end{equation}
$T_{VD}$ and $T_{IS}$ denote sets of testing samples where the ground truth needs-meeting way is via delivery and in store, respectively. In the following, we will use these metrics to measure the living needs prediction performance of our system and baseline systems.
\subsection{Baselines}\label{sec::baselines}
To illustrate the effectiveness of our system, we compare it with two baselines widely in actual production environments, including \textbf{DIN~\cite{zhou2018deep}}, \textbf{DNN~\cite{cheng2016wide}}, \textbf{DCN~\cite{wang2017deep}}, 
\textbf{ESMM~\cite{ma2018entire}},
and \textbf{MMOE~\cite{ma2018modeling}}. We will provide a detailed description of these baselines in the appendix. DIN is a recommendation algorithm that leverages deep neural networks to analyze users' historical behavior and make predictions about their potential interests. It uses an attention-based mechanism to weigh the importance of different historical behaviors for predicting the current interest of a user. As for DNN, We follow the design of Wide \& Deep learning to build a Deep Neural Network (DNN) based system for our task. It can learn both simple and complex relationships in the data. DCN is proposed to keep the benefits of a DNN model while introducing a cross network that is more efficient in learning certain bounded-degree feature interactions. It applies feature crossing at each layer, and it doesn't require manual feature engineering, adding minimal extra complexity to the DNN model. ESMM estimates post-click conversion rates for recommendation systems by using sequential user actions and a feature representation transfer learning strategy to alleviate sample selection bias and data sparsity. MMoE is a multi-task learning approach that learns to model task relationships from data. It adapts the Mixture-of-Experts (MoE) structure to multi-task learning by sharing the expert submodels across all tasks, while also having a gating network trained to optimize each task. 
For a clear comparison, the input features for all baselines are kept the same as those for our model.
\subsection{Additional Ablation Study}
To gain a deeper understanding of the impact of each component design of our model, we also conduct ablation studies with a particular focus on the effects of the group behavior pattern features and the spatiotemporal context features. When removing the group behavior pattern features from the model input, the performance decreased by 1.06\%. When removing the spatiotemporal context features from the model input, the performance decreased by 1.46\%. These results provide compelling evidence of the pivotal role that these two features play in accurately predicting users' daily living needs.

\end{document}